\documentclass[twocolumn,superscriptaddress,longbibliography,aps,prx,
preprintnumbers]{revtex4-2}

\usepackage[normalem]{ulem}
\usepackage{graphicx}
\usepackage{bm}
\usepackage{color}
\usepackage{epstopdf}
\usepackage{amsmath}
\usepackage{amssymb}
\usepackage{epstopdf}
\usepackage{lipsum}

\usepackage{gensymb}

\usepackage{multirow}

\usepackage{scrextend}

\usepackage[urlcolor=blue,colorlinks=true,citecolor=blue,linkcolor=blue,pdfstartview={FitH},bookmarks=false]{hyperref}

\usepackage[dvipsnames]{xcolor}
\definecolor{mygreen}{rgb}{0.0, 0.6, 0.0}
\definecolor{pjorange}{rgb}{0.8, 0.3, 0.0}
\definecolor{jlblue}{rgb}{0.2, 0.5, 0.7}

\graphicspath{{fig/}{./fig/}{.}}

\begin{document}

\title{
Phononic drumhead surface state in distorted kagome compound RhPb
}

\author{Andrzej~Ptok}
\email[e-mail: ]{aptok@mmj.pl}
\affiliation{\mbox{Institute of Nuclear Physics, Polish Academy of Sciences, W. E. Radzikowskiego 152, PL-31342 Krak\'{o}w, Poland}}

\author{William~R.~Meier}
\email[e-mail: ]{javamocham@gmail.com}
\altaffiliation[Current address: ]{Materials Science \& Engineering Department, University of Tennessee Knoxville, Knoxville, Tennessee 37996, USA}%
\affiliation{Material Science \& Technology Division, Oak Ridge National Laboratory, Oak Ridge, Tennessee 37831, USA}%

\author{Aksel~Kobia\l{}ka}
\affiliation{Department of Physics, University of Basel, Klingelbergstrasse 82, CH-4056 Basel, Switzerland}

\author{Surajit Basak}
\affiliation{\mbox{Institute of Nuclear Physics, Polish Academy of Sciences, W. E. Radzikowskiego 152, PL-31342 Krak\'{o}w, Poland}}

\author{Ma\l{}gorzata~Sternik}
\affiliation{\mbox{Institute of Nuclear Physics, Polish Academy of Sciences, W. E. Radzikowskiego 152, PL-31342 Krak\'{o}w, Poland}}

\author{Jan~\L{}a\.{z}ewski}
\affiliation{\mbox{Institute of Nuclear Physics, Polish Academy of Sciences, W. E. Radzikowskiego 152, PL-31342 Krak\'{o}w, Poland}}

\author{Pawe\l{}~T.~Jochym}
\affiliation{\mbox{Institute of Nuclear Physics, Polish Academy of Sciences, W. E. Radzikowskiego 152, PL-31342 Krak\'{o}w, Poland}}

\author{Michael~A.~McGuire}
\affiliation{Material Science \& Technology Division, Oak Ridge National Laboratory, Oak Ridge, Tennessee 37831, USA}%

\author{Brian~C.~Sales}
\affiliation{Material Science \& Technology Division, Oak Ridge National Laboratory, Oak Ridge, Tennessee 37831, USA}%

\author{Hu~Miao}
\affiliation{Material Science \& Technology Division, Oak Ridge National Laboratory, Oak Ridge, Tennessee 37831, USA}%

\author{Przemys\l{}aw~Piekarz}
\affiliation{\mbox{Institute of Nuclear Physics, Polish Academy of Sciences, W. E. Radzikowskiego 152, PL-31342 Krak\'{o}w, Poland}}

\author{Andrzej~M.~Ole\'{s}}
\email[e-mail: ]{a.m.oles@fkf.mpg.de}
\affiliation{Max Planck Institute for Solid State Research,
Heisenbergstrasse 1, D-70569 Stuttgart, Germany}
\affiliation{\mbox{Institute of Theoretical Physics, Jagiellonian University,
Profesora Stanis\l{}awa \L{}ojasiewicza 11, PL-30348 Krak\'{o}w, Poland}}

\date{\today}

\begin{abstract}
RhPb was initially recognized as one of a CoSn-like compounds with $P6/mmm$ symmetry, containing an ideal kagome lattice of $d$-block atoms.
However, theoretical calculations predict the realization of the phonon soft mode which leads to the kagome lattice distortion and stabilization of the structure with $P\bar{6}2m$ symmetry [A. Ptok {\it et al.}, \href{https://link.aps.org/doi/10.1103/PhysRevB.104.054305}{Phys. Rev. B {\bf 104}, 054305 (2021)}].
Here, we present the single crystal x-ray diffraction results supporting this prediction.
Furthermore, we discuss the main dynamical properties of RhPb with $P\bar{6}2m$ symmetry.
The bulk phononic dispersion curves contain several flattened bands, Dirac nodal lines, and triple degenerate Dirac points.
As a consequence, the phononic drumhead surface state is realized for the (100) surface, terminated by the zigzag-like edge of Pb honeycomb sublattice.
\end{abstract}

\maketitle

\section{Introduction}
\label{sec.intro}

Discovery of the topological insulators with conducting surface states in the form of the Dirac cone~\cite{zhang.liu.09,xia.qian.09,hsieh.xia.09,alpichshev.analytis.10} opened a period of the intensive studies in the subject of fermionic topological systems~\cite{hasan.kane.10,qi.zhang.11,armitaga.male.18}.
However, a realization of the non-trivial topological states is not limited only to fermionic systems but can be also expected in bosonic ones as well~\cite{lu.joannopoulos.14,yang.gao.15,ozawa.price.19,liu.chen.20,mcclarty.22}.
We can find several examples of the realization of phonon Dirac/Weyl points~\cite{miao.zhang.18,li.xie.20,li.xie.18,wang.xia.20,chen.wang.21,zhang.song.18,zhong.liu.21,xie.liu.21,liu.li.21,liu.hou.19,liu.qian.20,liu.wang.21b,wang.yuan.22,ding.wang.23,yang.xie.23}, nodal lines~\cite{wang.yang.22,liu.jin.21,xie.yuan.21,wang.wang.22,basak.kobialka.23,zhou.zhang.21}, nodal rings~\cite{zheng.xia.20,jin.chen.18,zheng.zhan.21,wang.chen.21,zhou.zhang.21,chen.xie.22}, and nodal nets~\cite{chen.huang.21,zhou.chen.21,ding.sun.22,zhu.wu.22}.
Topological properties are also manifested by the emergence of phonon surface states~\cite{li.xie.18,li.wang.20,wang.xia.20,wang.yuan.21,jin.chen.18,zheng.zhan.21,liu.wang.21,wang.yuan.21b,chen.wang.21,zhang.song.18,zhong.liu.21,liu.fu.21,wang.zhou.21,basak.kobialka.23,yang.wang.23} or phonon Hall effect~\cite{strohm.rikken.05,sheng.sheng.06,kagan.maksimov.08,zhang.jie.10,qin.zhou.12,saito.misaki.19,zhang.zhang.19}.
As a result, existence of the topological phonons with non-zero Berry curvature~\cite{zhang.niu.15} can give rise to development of nanodevices based on the heat transfer manipulation, i.e., in {\it phononics}~\cite{li.ren.12}.

In the context of the topological properties, the kagome-supported systems have been of great interest recently.
The basic property of the system with the kagome lattice is the formation of flat electronic bands~\cite{lin.choi.18,yin.zhang.19,li.wang.21}.
One such example of the kagome system are CoSn-like compounds, which combine kagome and honeycomb layers~\cite{sales.yan.19,meier.du.20,liu.li.20,huang.zheng.22}.
The electronic band structure exhibits a flat band and Dirac fermions~\cite{sales.yan.19,meier.du.20,kang.fang.20,kang.ye.20,lin.wang.20,huang.zheng.22,liu.li.20,han.inoue.21,sales.meier.21}.
Some CoSn-like compounds exhibit frustrated magnetism with an important role for itinerant electrons~\cite{xie.chen.21}, like magnetically ordered FeGe~\cite{zeng.kent.06}, or FeSn~\cite{sales.yan.19,kakihana.nishimura.19,khadka.thapaliya.20,sales.meier.21}.
However, most commonly, these compounds are paramagnets, such as CoSn~\cite{kakihana.nishimura.19}.

\begin{figure}[!b]
\centering
\includegraphics[width=\columnwidth]{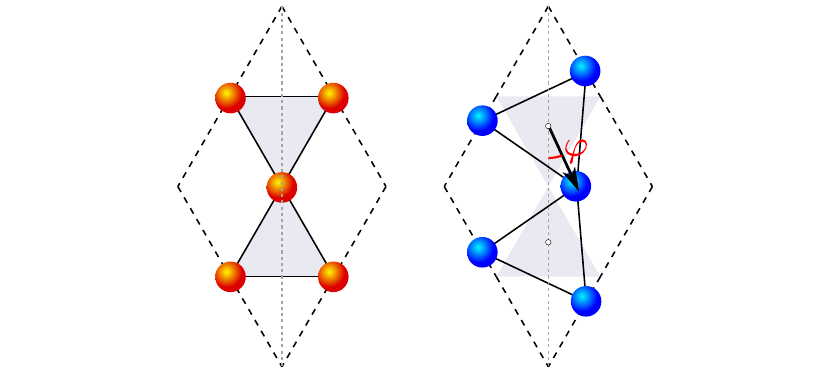}
\caption{
Schematic representation of the {\it ideal} (left panel) and {\it distorted} (right panel) kagome lattices, realized in the system with symmetries $P6/mmm$ and $P\bar{6}2m$, respectively.
In the ideal kagome lattice, the system is centrosymmetric.
The rotation of perfectly ordered triangles by angle  $\varphi$ transforms the system 
into the non-centrosymmetric structure.
\label{fig.schem}
}
\end{figure}

\paragraph*{Motivation:}
Typically kagome layer in CoSn-like compounds crystallizes with the $P6/mmm$ symmetry~\cite{sales.yan.19,meier.du.20}.
In such a structure, the $d$-block element (e.g. Fe, Co, Ni, Rh, or Pt) forms an ideal kagome sublattice, while the $p$-block element (e.g. Ge, In, Sn, Tl, or Pb) has two non-equivalent positions: one of the $p$-block elements is located in the plane of the kagome sublattice when the second's position forms honeycomb sublattice intercalated between two kagome-sublattice planes.
RhPb satisfies the mentioned conditions, and therefore it should behave as an ideal kagome metal~\cite{meier.du.20}.
However, theoretical investigations of the lattice dynamics of RhPb show that this compound should be unstable with $P6/mmm$ symmetry~\cite{ptok.kobialka.21}.
Stabilization of RhPb can be achieved by distorting the kagome lattice, resulting in $P\bar{6}2m$ symmetry (see Fig.~\ref{fig.schem}).
In this paper, we present experimental evidence for the distorted kagome structure in the RhPb system.
Furthermore, we investigate the dynamic properties of RhPb with distorted kagome lattice.
We show that this compound is an excellent candidate for the study of the phonon drumhead surface state.

The paper is organized as follows.
Firstly, we discuss the results indicating formation of the distorted kagome lattice in RhPb with $P\bar{6}2m$ symmetry (Sec.~\ref{sec.crys}).
Next, we discuss the theoretical study of the dynamical properties of this compound (Sec.~\ref{sec.phon}).
Finally, we summarize the paper with the main conclusions in Sec.~\ref{sec.sum}.


\section{Crystal structure}
\label{sec.crys}

\begin{figure}[!b]
\centering
\includegraphics[width=\linewidth]{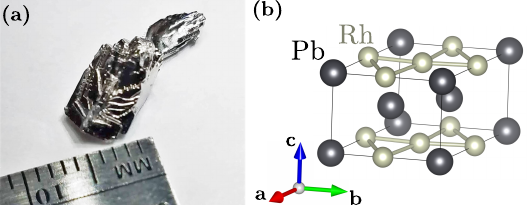}
\caption{Single crystal of RhPb grown from lead melt (a), and crystal structure of RhPb with $P\bar{6}2m$ symmetry (b). 
\label{fig.sample}
}
\end{figure}

Crystals of RhPb were grown from a high-temperature Pb-rich melt~\cite{meier.du.20}. 
A 1:3 atomic ratio of rhodium sponge (Alfa Aesar $99.95$\%) and lead slugs (Alfa Aesar Puratronic $99.999$\%) were loaded into one side of $2$~mL alumina Canfield crucible set~\cite{canfield.kong.16} and then sealed in a fused silica ampoule under vacuum with a hydrogen-oxygen torch.
The ampoule was placed in a box furnace and heated to $1000^{\circ}$C or $1100^{\circ}$C over $6$~h.
This temperature was held for 2 hours to dissolve the rhodium in lead and homogenize the fluid. 
The furnace was quickly cooled to $900^{\circ}$C over $3.5$~h and held for $0.5$~h before cooling to $750^{\circ}$C over $320$~h ($-0.47^{\circ}$C/h) to slowly precipitate the crystal. 
The hot ampule was removed from the furnace and inverted into a centrifuge to fling the remaining liquid of the crystals.
In the batch used for single crystal diffraction, $5.4$~g of reactants yielded a single $4 \times 7$~mm crystals weighing about $1.2$~g.
It had a slightly-skeletal hexagonal-prismatic shape with a shiny metallic luster [see Fig.~\ref{fig.sample}(a)]. 
Other batches of crystals yielded faceted euhedral blocky hexagonal prisms. 
Crystals of RhPb are brittle with a conchoidal fracture and have weak (001)-cleavage.
Broken surfaces sometimes reveal inclusions of bluish metallic Pb metal that contrast with the silver metallic RhPb surfaces.

To determine the crystal structure of RhPb, fragments of the $1.2$~g crystal were selected for single crystal X-ray diffraction (XRD). 
The crystal, approximately $60 \times 40 \times 10$~$\mu$m$^{3}$, was mounted on the end of a Kapton loop with Locktite glue for data collection at room temperature using a Bruker D8 Quest diffractometer ($0.71073$~\AA\ Mo Ka radiation). 
Data were collected, reduced, and analyzed using APEX3 software, including a semi-empirical absorption correction based on equivalent reflections. 
Structure refinement was performed using JANA 2020~\cite{petricek.dusek.14} for both $P6/mmm$ (space group No. 191) and $P\bar{6}2m$ (space group No. 189) symmetries.

\begin{table}[!t]
\caption{
\label{tab.StucturesBrief}
An important characteristic of refined structures of RhPb at 293 K from single crystal x-ray diffraction. Note the significant improvement of fit quality in the $P\bar{6}2m$ refinement.
}
\begin{ruledtabular}
\begin{tabular}{lcc}
Space Group & $P6/mmm$ & $P\bar{6}2m$ \\
\hline
$x_\text{Rh}$ & $1/2$ & $0.4775(2)$ \\
Twin fraction & N/A & $0.568(25)$ \\
Triangle rotation, $\varphi$ in Fig.~\ref{fig.schem}& $0^{\circ}$ & $4.45(3)^{\circ}$ \\
\hline
R(obs) (\%)	& $6.01$ & $2.83$ \\
Goodness of fit obs & $4.91$ & $1.65$ \\
\end{tabular}
\end{ruledtabular}
\end{table}

\begin{table}[!b]
\caption{
Comparison of the experimental and theoretical lattice constants for RbPb with different symmetries.
}
\begin{ruledtabular}
\begin{tabular}{lcc}
 & $a$ (\AA) & $c$ (\AA) \\
 \hline
Exp. $293(2)$~K & $5.6794(4)$ & $4.4311(3)$ \\
Exp. $15$~K (Ref.~\cite{meier.du.20}) & $5.66601(2)$ & $4.41267(1)$ \\
$P6/mmm$ & $5.740$ & $4.487$ \\
$P\bar{6}2m$ & $5.762$ & $4.466$ \\
\end{tabular}
\end{ruledtabular}
\label{tab.latt}
\end{table}

The first-principles density functional theory (DFT) calculations were performed using the projector augmented-wave (PAW) potentials~\cite{blochl.94} implemented in the Vienna Ab initio Simulation Package ({\sc Vasp}) code~\cite{kresse.hafner.94,kresse.furthmuller.96,kresse.joubert.99}.
Calculations were made within the generalized gradient approximation (GGA) in the Perdew, Burke, and Ernzerhof (PBE) parameterization~\cite{pardew.burke.96}.
The energy cutoff for the plane-wave expansion was set to $350$~eV.
Optimizations of structural parameters (lattice constants and atomic positions) are performed in the primitive unit cell using the $10 \times 10 \times 6\, 
{\bm k}$--point grid in the Monkhorst--Pack scheme~\cite{monkhorst.pack.76}
As a break condition of the optimization loop, we take the energy difference of $10^{-6}$~eV and $10^{-8}$~eV for ionic and electronic degrees of freedom, respectively.

\begin{figure*}[!t]
\centering
\includegraphics[width=0.85\linewidth]{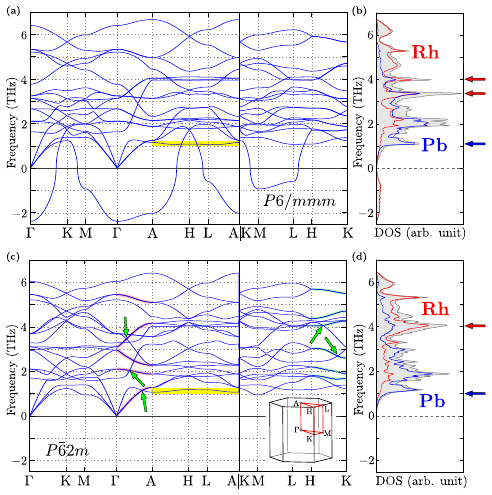}
\caption{
Phonon dispersion relations (left column) and phonon density of states (right column) for RhPb with $P6/mmm$ (top row) and $P\bar{6}2m$ (bottom row) symmetry.
The marked red and green branches correspond to the double degenerate band along the $\Gamma$-A and K-H directions, respectively.
Color lines in the right panels are related to the total DOS (gray) or partial DOS (red and blue for lighter Rh and heavier Pb atoms, respectively).
The inset in panel (c) represents the Brillouin zone and its high symmetry points.
\label{fig.phon2}
}
\end{figure*}

\paragraph*{Structure:}
Initially, the RhPb structure was assumed to be possess $P6/mmm$ symmetry~\cite{meier.du.20}. 
However, theoretical analysis of the RhPb dynamical properties show that such a system is unstable~\cite{ptok.kobialka.21}.
Our single crystal refinement do indeed reveal that RhPb adopts the distorted $P\bar{6}2m$ structure based on a distinctly better fit quality over the $P6/mmm$ solution (Tab.~\ref{tab.StucturesBrief}). 
DFT calculations estimate very subtle differences in the lattice parameters for the two structures (Tab.~\ref{tab.latt}) but more stable phonons in the $P\bar{6}2m$ structure (detailed discussion in Sec.~\ref{sec.phon_disp}).

In the case of $P6/mmm$ symmetry, the Rh atoms are located in the Wyckoff position $3f$ ($1/2$,$1/2$,$0$), while the Pb atoms in two non-equivalent Wyckoff positions $2d$ ($1/3$,$2/3$,$1/2$) and $1a$ ($0$,$0$,$0$).
In the case of the $P\bar{6}2m$ symmetry [see Fig.~\ref{fig.sample}(b)], the $3f$ position of the Rh atom is calculated to lie at ($0.467$,$0.467$,$0$).
Experimentally, the position of the Rh atom was estimated as ($0.4775(2)$,$0.4775(2)$,$0$).
The distortion introduced to the system corresponds to the rotation of the triangles that form an ideal kagome lattice with angle $\varphi = 6.5^{\circ}$ determined by DFT, while the experimentally obtained value is $4.46^{\circ}$ (see Fig.~\ref{fig.schem}). 
The rotation of the kagome triangles in RhPb mirrors the modified kagome pattern seen in the ZrNiAl-type materials~\cite{pott.schefzyk.81,kim.echizen.03,morosan.budko.04,matar.riecken.07,deboer.elenbaas.08,qian.nie.19,su.shang.21,asaba.ivanov.21,asaba.ivanov.21,kumar.luo.22}.
More crystallographic data (e.g. CIF files as well as XRD refinements) can be found in the Supplemental Material (SM)~\footnote{Supplemental Material at [URL will be inserted by publisher] for more details of the crystallographic data, XRD refinements, and additional numerical results. This Supplemental Material contain Ref.~\cite{sales.yan.19,meier.du.20,liu.li.20,huang.zheng.22,kang.fang.20,meier.du.20,jovanovic.schoop.22,zeng.kent.06,sales.yan.19,kakihana.nishimura.19,khadka.thapaliya.20,sales.meier.21,herman.kuglin.63,shanavas.popovic.14}.}.

Regardless of the symmetry, RhPb exhibits a metallic band structure (see Fig.~\ref{fig.el_band} in SM~\cite{Note1}). 
The electronic band structure obtained for both symmetries is very similar, and in correspondence to other CoSn-like compounds, it contains several flattened bands (detailed discussion can be found in Sec.~\ref{sec_sm.ele} in the SM~\cite{Note1}.


\section{Dynamical properties}
\label{sec.phon}


The dynamical properties were calculated using the direct {\it Parlinski–Li–Kawazoe} method~\cite{parlinski.li.97}, implemented in {\sc Phonopy} package~\cite{togo.tanaka.15}. 
Within this method, the interatomic force constants (IFC) are calculated from the Hellmann-Feynman (HF) forces acting on the atoms after displacements of individual atoms inside the supercell.
We performed these calculations using the $2 \times 2 \times 2$ supercell with $48$ atoms, and the reduced {\bf k}-grid $3 \times 3 \times 3$.
Next, the IFC were used to study the surface states, by calculations of the surface Green's function for semi-infinite system~\cite{sancho.sancho.85}, by {\sc WannierTools}~\cite{wu.zhang.18}.
Additionally, we also calculate the phonon dispersion curves for a finite temperature. 
In this case, the calculations were performed for the thermal distribution of multi-displacement of atoms~\cite{hellman.abrikosov.11}, generated within the {\sc hecss} procedure~\cite{jochym.lazewski.21}.
The total energy and HF forces acting on all atoms are calculated with {\sc Vasp} for 100 different configurations of atomic displacements in the supercell.
In dynamical properties calculations, we include second- and third-order phonon contributions, which correspond to the harmonic and cubic IFC, respectively.

\subsection{Phonon dispersion curves}
\label{sec.phon_disp}

The phonon dispersion relations and the phonon density of states (DOS) for RhPb with both symmetries are presented in Fig.~\ref{fig.phon2}.
In the case of the $P6/mmm$ symmetry, there exists the imaginary soft mode (presented as negative frequencies) [see Fig.~\ref{fig.phon2}(a)].
This soft mode, with frequency of 
$-2.59$~THz at the $\Gamma$ point, is characterized by the $B_\text{1u}$ symmetry.
Atomic displacements induced by this 
mode lead to the rotation of the triangles forming the ideal kagome lattice~\cite{ptok.kobialka.21} (see Fig.~\ref{fig.schem}).
As a consequence, the $P\bar{6}2m$ symmetry is stabilized -- after the transformation the phonon dispersion does not exhibit any 
imaginary modes, so
all frequencies are positive [Fig.~\ref{fig.phon2}(c)].
The analysis of the zone-center mode frequencies and symmetries shows 
that in the distorted structure the mode corresponding to the soft-mode has a frequency of $3.2$~THz and an $A$-like symmetry.

Soft modes realized for the $P6/mmm$ symmetry are associated only with the Rh atoms vibration, which is reflected in the phonon density of states.
In practice, all spectral weights at DOS related to the soft mode [negative frequencies at Fig.~\ref{fig.phon2}(b)] correspond to the Rh atoms contribution.
For negative frequencies the contribution of Pb atoms is negligible.
Indeed, stabilization of RhPb with the P$\bar{6}$2m symmetry modified mainly the Rh contribution [cf.~Fig.~\ref{fig.phon2}(b) and \ref{fig.phon2}(d)].
As expected, independently of the system symmetry, the vibrations of the heavy Pb atoms are realized mostly in the lower frequencies range.
The vibrations of lighter Rh atoms exist in the higher frequency range.

Here, we should point out that the spontaneous kagome rotation can be also achieved in other
compounds, like e.g. MgCo$_{6}$Ge$_{6}$~\cite{sinha.vivanco.21} at $T=100$~K.
Nevertheless, in the case of RhPb, the theoretical investigation of temperature-dependent phonon dispersion for $P6/mmm$ always shows soft mode at the $\Gamma$ point (even up to $1500$~K).
From this, we can conclude that the RhPb compound crystallizes only with 
the $P\bar{6}2m$ symmetry, while the structure with the $P6/mmm$ symmetry is unstable even at high temperatures.

Symmetry realized by RhPb has an impact on the irreducible representation of phonons at the $\Gamma$ point. 
Indeed, the exact analysis presented in the Sec.~\ref{sec_sm.active} in SM~\cite{Note1} clearly shows differences between these two phases.
In fact, due to the different number of active modes in both symmetries, e.g. visible in the Raman spectroscopy, we can gain additional evidence for the formation of the $P\bar{6}2m$ crystal symmetry.


\subsection{Phonon flat bands}

\begin{figure}[!b]
\centering
\includegraphics[width=0.95\linewidth]{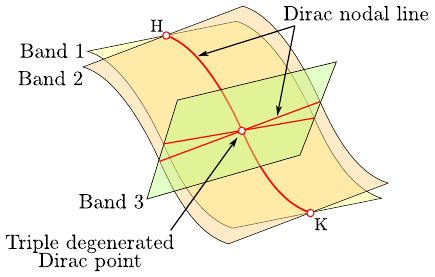}
\caption{
Schematic representation of triple degenerated Dirac point forming along the K--H path.
\label{fig.ph_dirac}
}
\end{figure}

Typically, the kagome lattice allows for the realization of the flat bands -- this property is observed in the electronic band structure of CoSn-like compounds~\cite{kang.fang.20,liu.li.20,huang.zheng.22,meier.du.20}.
However, similar behavior is also observed in phonon dispersion curves.
For both symmetries, there are several flat bands with a weak dispersion along the $\Gamma$--K--M--$\Gamma$ and A--H--L--A paths [e.g. branch marked by yellow areas in Figs.~\ref{fig.phon2}(a) and \ref{fig.phon2}(c)].
A previous study of the similar CoSn system, suggests that the flat bands at low frequencies should be related with the collective vibrations of $d$-block atoms (i.e., Rh in our case) within the (ideal or distorted) kagome lattice~\cite{yin.shumiya.20}.
However, exact analysis of the phonon DOS clearly shows that these modes are related to the vibrations of Pb atoms [marked by blue arrows in Figs.~\ref{fig.phon2}(b) and \ref{fig.phon2}(d)].
Additionally, at this range of frequencies, the chiral phonons (i.e., circulations of the atoms around the equilibrium position) were predicted within the Pb honeycomb sublattice~\cite{ptok.kobialka.21}.

\begin{figure*}
\centering
\includegraphics[width=\linewidth]{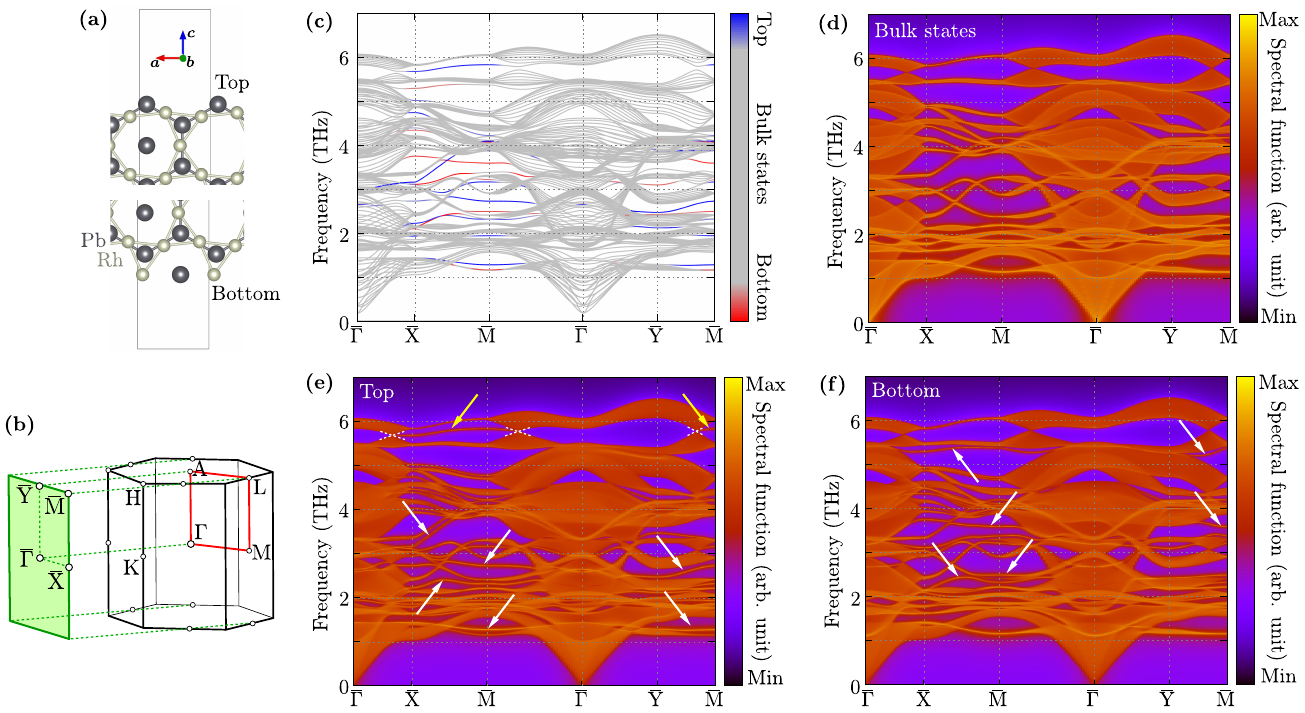}
\caption{
The phonon surface states realized at the edge of the (100) surfaces of RhPb with $P\bar{6}2m$ symmetry are presented in~(a).
For simplification, we show only bonding between Rh atoms forming a distorted kagome lattice and Pb atoms forming a honeycomb lattice. 
The Pb atom in the center of hexagons is placed in the plane of a distorted kagome lattice.
(b) Relation between the bulk (3D) and surface (2D) Brillouin zone.
Phonon band structure calculated for slab-like structure (c), containing ten layers.
The colors correspond to the states related to vibrations of the atoms at the top surface (blue), central bulk-like part (gray), or bottom surface (red).
The panels from (d) to (f) present the spectral function calculated for the central bulk-like part, the top surface, and the bottom surface part of the slab, respectively.
The arrows in (e) and (f) show the locations of the soft modes of the phonons realized on the top or bottom surfaces, respectively.
\label{fig.ph_ss}
}
\end{figure*}

The realization of mentioned flat bands is reflected in the phonon DOS, by the relatively sharp peaks [marked by red arrows in Figs.~\ref{fig.phon2}(b) and \ref{fig.phon2}(d)].
Moreover, the phonon DOS describes the frequency distribution of normal modes inside the
whole Brillouin zone, in contrast to the phonon dispersion curves which only represent 
the modes along high symmetry directions.
Indeed, the true flat-like bands 
can be realized in $P6/mmm$ symmetry, where the peaks in DOS are very sharp [see Fig.~\ref{fig.phon2}(b)].
Contrary to this, in the case of $P\bar{6}2m$ symmetry, the stronger 
${\bm k}$-dependent dispersion is uncovered by the existence of much broader peaks in phonon DOS [see Fig.~\ref{fig.phon2}(d)].
It means, that the phonon bands for the $P\bar{6}2m$ symmetry are more dispersive than in the case of the $P6/mmm$ symmetry.


\subsection{Bands degeneracy and Dirac points/lines}

\begin{figure*}
\centering
\includegraphics[width=\linewidth]{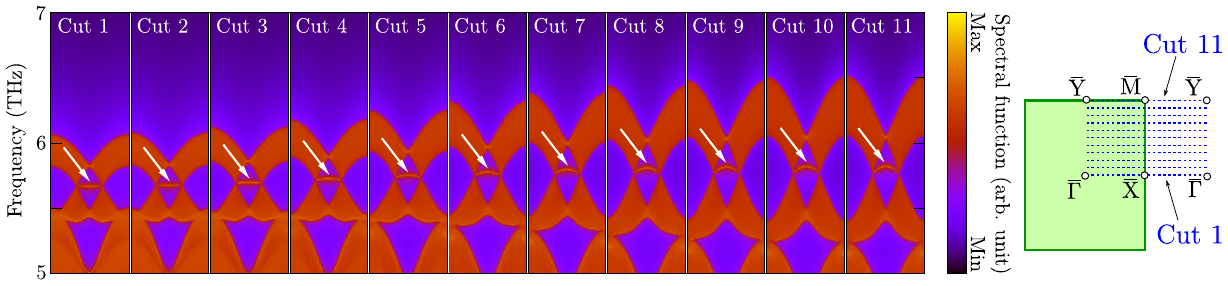}
\caption{
The surface spectral function for different cuts of the surface (2D) Brillouin zone (right panel), presenting ${\bm k}$-dependence of the phonon drumhead surface state (marked by white arrows).
\label{fig.ph_ss_cross}
}
\end{figure*}

The realization of three- and six-folded rotational symmetry allows preserving band degeneracy from $\Gamma$ points. 
Indeed, the degeneracy preservation is well visible along the $\Gamma$--A path (i.e., for $z$ direction, perpendicular to the honeycomb and kagome layers) [branches marked by the red solid line in Fig~\ref{fig.phon2}(c)].
Additionally, the hexagonal symmetry affects the band structure along the K--H path, where some bands are double-degenerated and form the Dirac nodal lines.
Irreducible representations at the K (H) point allow for the realization of double degenerated points with the K$_5$ and K$_6$ (H$_5$ and H$_6$) symmetry. However, the symmetry of this state is preserved for any point P with coordinates (1/3,1/3,u), within the double degenerated state 
with the P$_3$ symmetry.
Finally, the Dirac point at the K (H) point [visible e.g. in the form of characteristic band crossing around $5.6$~THz for both symmetries in Fig.~\ref{fig.phon2}(a) and (c)], is ``stretched'' out along all the K--H path forming the Dirac line.

Additionally, in the phonon dispersion curves a few crossings of the bands with different degeneracy along the $\Gamma$--Z and along the K--H directions can be found [places shown by green arrows in Fig.~\ref{fig.phon2}(c)].
This type of crossing forms the Dirac point with tripled degeneracy [see Fig.~\ref{fig.ph_dirac}].
In this case, two bands (No.~1 and 2) are crossing along the K--H path, as discussed in the previous paragraph.
Additional band (no.~3) crosses initial two bands creating additional Dirac nodal lines (by crossing one of them), and triple degenerated Dirac point (in places of crossing both bands \mbox{simultaneously}).
Here we would like to point out that the realization of (bulk) Dirac point and nodal lines has a strong effect on possible surface states. 
Indeed, in the next paragraph, we will focus on this aspect.


\subsection{Phonon surface states}

Some hexagonal lattices, like honeycomb~\cite{brey.fertig.06,wakabayashi.takane.07,yao.yang.09} and kagome~\cite{yang.nagaosa.14,li.zhuang.18} nets, can form the electronic surface states at the zigzag-like edge of the lattice.
This is also true in the case of the bosonic systems~\cite{mook.henk.14,malz.knolle.19,zhong.wang.19,li.wang.20,xi.ma.21}.
Indeed, the phonon zigzag edge modes could be realized in RhPb at the (100) surface [Fig.~\ref{fig.ph_ss}(a)].
Similar behavior was also earlier reported for NbReSi, which possesses the same symmetry~\cite{basak.ptok.23}.

The calculation of the phonon surface states of RhPb for the (100) surface is presented in Fig.~\ref{fig.ph_ss}.
In our calculations, we consider two types of terminations.
The zigzag edge of the Pb honeycomb lattice is realized in both of them.
The ``top'' surface contains also a chain of Rh atoms from the kagome lattice.
Similarly, the ``bottom'' surface contains a chain of Rh-Pb atoms from the triangular lattice formed by the Rh kagome net decorated by Pb atoms represented by non-bonded black atom in Fig.~\ref{fig.ph_ss}(a).
Slab-like calculations (for ten layers of RhPb) clearly show the existence of surface states, independently of the termination [blue and red lines in Fig.~\ref{fig.ph_ss}(c)].
The phonon dispersion curves in this case contain much more branches than the dispersion curves for the bulk [Fig.~\ref{fig.phon2}(c)], what is the consequence of projection of all phonon states from the bulk (3D) Brillouin zone onto the 2D Brillouin zone of the surface [see Fig.~\ref{fig.ph_ss}(b)].
It is interesting that, in the case of the surface directly terminated by the zigzag edge of the Pb honeycomb lattice (i.e., top surface), the phonon surface states at the highest frequencies exhibit behavior similar to this observed in electronic surface states at the zigzag edge in graphene nanoribbon~\cite{brey.fertig.06} [see the states with frequencies around $5.75$~THz, marked by yellow arrows in Fig.~\ref{fig.ph_ss}(e)].

\begin{figure}[!b]
\centering
\includegraphics[width=0.88\linewidth]{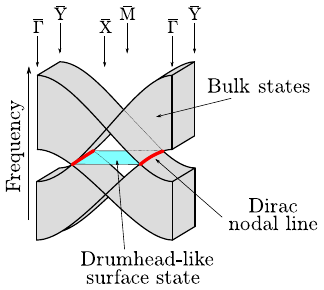}
\caption{
Schematic representation of the forming of phonon drumhead surface state.
\label{fig.drum}
}
\end{figure}

Direct calculations of the spectral functions for the bulk and surface regions are presented in \mbox{Figs.~\ref{fig.ph_ss}(d)--\ref{fig.ph_ss}(f)}.
The phonon surface states can be in a relatively simple way extracted from the bulk spectra, by comparing the spectral function related to the bulk states [see Fig.~\ref{fig.ph_ss}(d)] with the adequate spectral function for a specific surface [Fig.~\ref{fig.ph_ss}(e) or \ref{fig.ph_ss}(f)].
Additional states, i.e., surface states, are marked by arrows.
Comparing the frequencies of surface states with the bulk phonon DOS [Fig.~\ref{fig.phon2}(d)], we see that the surface states in the low (high) frequency range should be realized mostly by Pb (Rh) atoms.

Probably, the most interesting properties of the slab-like structure can be observed for the highest branches of the spectrum. 
First, the signature of the bulk Dirac ``point'' [i.e., Dirac lines from the K-H direction in the 3D bulk Brillouin zone are projected on the 2D surface Brillouin zone, see Fig.~\ref{fig.ph_ss}(b)] can be well recognized in the spectrum [white dashed crosses in Fig.~\ref{fig.ph_ss}(e)].
Between these points, the mentioned earlier zigzag-like edge mode is formed.

However, a more precise analysis of this latter state uncovers the realization of the drumhead phononic surface state.
Indeed, as we can recognize in Fig.~\ref{fig.ph_ss_cross}, these surface state is realized independently of $\bm k$, forming 2D surface state between the two bulk Dirac lines mentioned earlier.
Formation of the drumhead phonon surface state is presented schematically in Fig.~\ref{fig.drum}. 
Remarkably, the projected bulk phonon surface state create a graphene-like spectrum, independently of ${\bm k}$.
Note that the band crossing discussed earlier (red line in Fig.~\ref{fig.drum}) is visible as a Dirac ``point'' for any
${\bm k}$.
Finally, the drumhead surface state are realized between the Dirac nodal lines. 
Curiously, the frequency of phonon drumhead surface state strongly depends on $\bm k$ (Fig.~\ref{fig.ph_ss_cross}). 


\section{Summary}
\label{sec.sum}

In this paper we discuss the basic properties of RhPb with the distorted kagome lattice of Rh atoms.
Initially presented theoretical calculations~\cite{ptok.kobialka.21}, predict the realization of the distorted kagome lattice in RhPb.
Indeed, our single crystal diffraction results confirm the predicted distortion of the kagome net in RhPb. 

We presented a study of the dynamical properties of the bulk RhPb compound.
In such a system, the emergence of several flattened phonon bands is possible.
However, a more precise analysis shows that phonons in this band have even broader dispersion than for an ideal kagome lattice.
This behavior is clearly seen in the phonon density of states.

The phonon dispersion curves exhibit several interesting features, namely the (bulk) Dirac nodal lines and triple degenerated Dirac points. Such structures have consequences for the observed surface states.
The most prominent example is the realization of phonon drumhead surface state, between two (bulk) Dirac nodal lines projected on the surface Brillouin zone. 
In this context, the RhPb crystal with the distorted kagome lattice is an excellent platform to study the interplay between topological phonon surface states and electronic flat bands.

\begin{acknowledgments}
We kindly thank Mark O. Goerbig for insightful discussions.
Some figures in this work were rendered using {\sc Vesta}~\cite{momma.izumi.11} and {\sc XCrySDen}~\cite{kokalj.99} software.
A.P. is grateful to Laboratoire de Physique des Solides in Orsay (CNRS, University Paris Saclay) for hospitality during a part of the
work on this project.
Work by W.R.M., M.A.M, B.C.S., and H.M.~was funded by the U.S. Department of Energy, Office of Science, Basic Energy Sciences, Materials Sciences and Engineering Division. A.M.O. is grateful for support via the Alexander von Humboldt 
Foundation Fellowship (Humboldt-Forschungspreis).
We kindly acknowledge support by National Science Centre (NCN, Poland) 
under Project No.~2021/43/B/ST3/02166. 
\end{acknowledgments}


\bibliography{biblio.bib}


\clearpage
\newpage


\begin{center}
  \textbf{\Large Supplemental Material}\\[.3cm]
  \textbf{\large Phononic drumhead surface state \\[.2cm] in distorted kagome compound RhPb}\\[.3cm]
  Andrzej Ptok {\it et al.}\\[.2cm]
(Dated: \today)
\\[1cm]
\end{center}

\setcounter{equation}{0}
\renewcommand{\theequation}{S\arabic{equation}}
\setcounter{figure}{0}
\renewcommand{\thefigure}{S\arabic{figure}}
\setcounter{section}{0}
\renewcommand{\thesection}{S\arabic{section}}
\setcounter{table}{0}
\renewcommand{\thetable}{S\arabic{table}}
\setcounter{page}{1}

\begin{figure*}
\centering
\includegraphics[width=\linewidth]{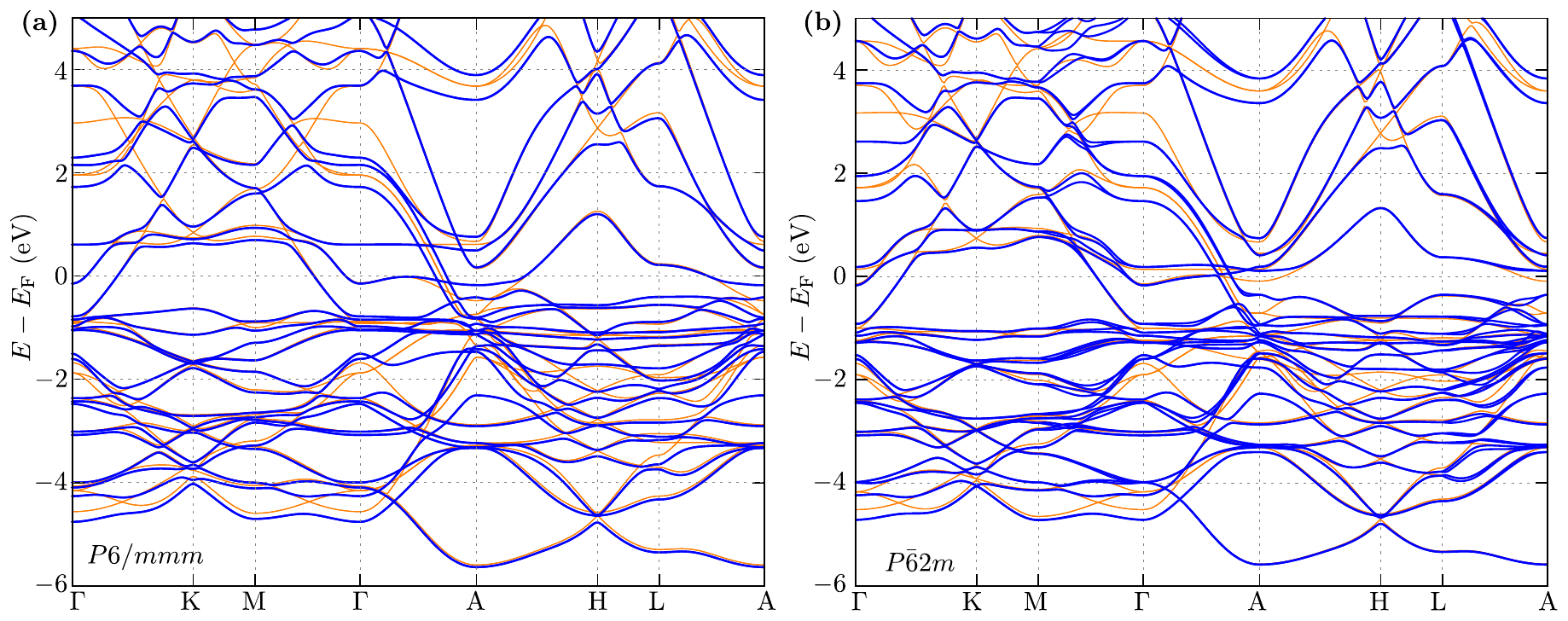}
\caption{
Electronic band structure of RhPb for $P6/mmm$ (a) and $P\bar{6}2m$ (b) symmetry.
Results in the presence (absence) of the spin--orbit coupling, are shown by a blue (orange) line.
\label{fig.el_band}
}
\end{figure*}
\begin{figure}[!b]
\centering
\includegraphics[width=0.95\linewidth]{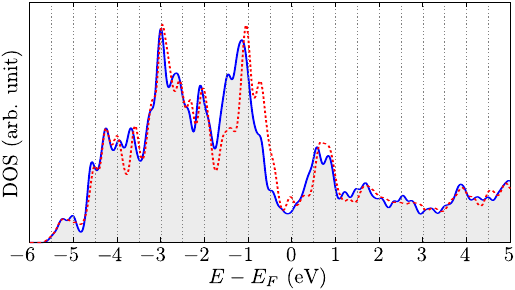}
\caption{
Comparison of the electronic density of states of RhPb with $P6/mmm$ and $P\bar{6}2m$ symmetries (red and blue lines, respectively).
\label{fig.el_dos}
}
\end{figure}

In this Supplemental Material, we present additional results:
\begin{itemize}
\item Table~\ref{tab.exp} -- Crystallographic data obtained from x-ray diffraction.
\item Table~\ref{tab.atom_disp} -- Atomic displacement parameters. 
\item {\bf Section~\ref{sec_sm.ele}} -- Section describing electronic band structure of RhPb with $P6/mmm$ and $P\bar{6}2m$ symmetries.
\item ~~~~~Fig.~\ref{fig.el_band} -- Comparison of electronic band structure of RhPb with $P6/mmm$ and $P\bar{6}2m$ symmetries.
\item ~~~~Fig.~\ref{fig.el_dos} -- Comparison of the electronic density of states for RhPb with $P6/mmm$ and $P\bar{6}2m$ symmetries.
\item ~~~~Fig.~\ref{fig.el_fs} -- Fermi surface for RhPb with $P\bar{6}2m$.
\item ~~~~Fig.~\ref{fig.arpes} -- ARPES results for (001) surface.
\item {\bf Section~\ref{sec_sm.active}} -- Section discussing the phonon irreducible representation at $\Gamma$ point of RhPb with $P6/mmm$ and $P\bar{6}2m$ symmetries.
\item ~~~~Table~\ref{tab.freq} -- Characteristic frequencies and symmetries of the phonon modes at $\Gamma$ point of RhPb with $P6/mmm$ and $P\bar{6}2m$ symmetries.
\end{itemize}

\begin{table}[!t]
\caption{
\label{tab.exp}
Refined RhPb structures from single crystal x-ray diffraction. Note that the $P\bar{6}2m$ space group gives a better refinement.
}
\begin{ruledtabular}
\begin{tabular}{lcc}
Chem formula & \multicolumn{2}{c}{RhPb} \\
Formula wt. (g/mol$\cdot$f.u.) & \multicolumn{2}{c}{310.11} \\
Wavelength & \multicolumn{2}{c}{Mo K$_\alpha$ (0.71073~\AA)} \\
Temperature (K) & \multicolumn{2}{c}{293(2)} \\
\hline
Symmetry & $P6/mmm$ & $P\bar{6}2m$ \\
\hline
Z & 3 & 3 \\
Calc. density (g/cm$^3$) & $12.4801$ & $12.4801$ \\
$a$, $b$ (\AA) & $5.6794(4)$ & $5.6794(4)$ \\
$c$ (\AA) & $4.4311(3)$ & $4.4311(3)$ \\
$V$ (\AA$^3$) & $123.779(19)$ & $12.779(19)$ \\
$x_\text{Rh}$ & $1/2$ & $0.4775(2)$ \\
Twin fraction & N.A. & $0.568(25)$ \\
Rh hexagon angles ($^{\circ}$) & $120$ & $128.91(4)$ \\
 & & /$111.09(4)$\footnote{Triangles rotated by $4.45(3)^{\circ}$.} \\
\hline
F(000) & $381$ & $381$ \\
No. of reflections & $163$ & $264$ \\
No. of observed reflections	& $160$ & $259$ \\
Refined based on $|F|$ & & \\
Criteria for observed reflections & $I > 3 s(I)'$ & $I > 3 s(I)'$ \\
Overall completeness & & \\
Redundance index ranges & & \\
\multicolumn{3}{c}{\begin{tabular}{p{4cm}p{1.25cm}p{1cm}p{1.25cm}p{1cm}}
    $h$ & $-9$ & $9$ & $-9$ & $9$ \\ 
    $k$ & $-10$ & $4$ & $-10$ & $4$ \\ 
    $l$ & $-7$ & $7$ & $-7$ & $7$ \\
    \end{tabular}
} \\
No. of variables & 10 & 12 \\
N$_{ref}$/N$_{var}$ & & \\
R(obs) (\%)	& $6.01$ & $2.83$ \\
wR(obs) (\%) & $8.94$ & $3.55$ \\
R(all) (\%)	& $6,08$ & $4.23$ \\
wR(all) (\%) & $8.95$ & $4.04$ \\
Goodness of fit obs & $4.91$ & $1.65$ \\
Goodness of fit all & $4.86$ & $1.86$ \\
$\mu$ & $111.222$ & $111.222$ \\
Extinction coeff. & $1700(400)$ & $660(90)$
\end{tabular}
\end{ruledtabular}
\end{table}

\begin{table*}
\caption{
\label{tab.atom_disp}
Atomic displacement parameters.}
\begin{ruledtabular}
\begin{tabular}{cccccccc}
{$P6/mmm$} & $U_{11}$ & $U_{22}$ & $U_{33}$ & $U_{12}$ & $U_{13}$ & $U_{23}$ & $U_\text{iso}$ \\
Pb & $0.0096(6)$ & $0.0096(6)$ & $0.0058(9)$ & $0.0048(3)$ & $0$ & $0$ & $0.0084(5)$ \\
Pb & $0.0137(7)$ & $0.0137(7)$ & $0.0167(12)$ & $0.0068(4)$ & $0$ & $0$ & $0.0147(6)$ \\
Rh & $0.0256(12)$ & $0.0047(9)$ & $0.0082(13)$ & $0.0024(5)$ & $0$ & $0$ & $0.0152(8)$ \\
\hline
{$P\bar{6}2m$} & $U_{11}$ & $U_{22}$ & $U_{33}$ & $U_{12}$ & $U_{13}$ & $U_{23}$ & $U_\text{iso}$ \\
Pb & $0.0103(2)$ & $0.0103(2)$ & $0.0035(3)$ & $0.00517(11)$ & $0$ & $0$ & $0.00806(19)$ \\
Pb & $0.0092(3)$ & $0.0092(3)$ & $0.0157(5)$ & $0.00462(13)$ & $0$ & $0$ & $0.0114(2)$ \\
Rh & $0.0105(4)$ & $0.0069(4)$ & $0.0063(5)$ & $0.00345(19)$ & $0$ & $0$ & $0.0083(3)$
\end{tabular}
\end{ruledtabular}
\end{table*}

\section{Electronic properties}
\label{sec_sm.ele}

Typically, CoSn-like compounds, containing ideal kagome lattice of the {\it d}-block elements (i.e., P6/mmm symmetry), are characterized by the electronic flat-bands~\cite{sales.yan.19,meier.du.20,liu.li.20,huang.zheng.22,kang.fang.20}.
Similarly, to the case of the phonon dispersion, the electronic band structure possesses several nearly-flat bands.
The orbital character of the flat bands, with bands in range of around $0.5$~eV, are associated with the combination of $d_{xy}$ and $d_{x^{2}-y^{2}}$, or $d_{xz}$ and $d_{yz}$ orbitals~\cite{meier.du.20}.
In fact, the RhPb with P6/mmm symmetry exhibits the complex electronic band structure of the kagome lattice, as a consequence of the emergence of the isolated kagome net by Rh atoms~\cite{jovanovic.schoop.22}.
In this case, the kagome-like band structure is formed in energies between $-3$ and $-1$~eV [see Fig.~\ref{fig.el_band}(a)].
The introduction of the distortion in the kagome lattice does not change this feature [see Fig.~\ref{fig.el_band}(b)].

Unlike the ferromagnetic FeGe~\cite{zeng.kent.06} or antifferomagnetic FeSn~\cite{sales.yan.19,kakihana.nishimura.19,khadka.thapaliya.20,sales.meier.21}, RhPb does not exhibit magnetic order.
The electronic band structure decoupling is introduced by the relatively strong spin--orbit coupling (cf. band structures in the absence and presence of the spin--orbit coupling, presented by the orange and blue lines, respectively, in Fig.~\ref{fig.el_band}).
In practice, the strength of the spin--orbit coupling is strictly related to the atomic mass of the elements~\cite{herman.kuglin.63,shanavas.popovic.14}.
Indeed, in our case, the relatively weak impact of the spin-orbit coupling is observed in the range energies related to the $d$ orbitals of Rb, while stronger in the case of Pb atoms contribution.
For example, the band splitting along the $\Gamma$--A path is in the range $0.75$~eV (cf. orange and blue lines in Fig.~\ref{fig.el_band}).
Bigger splinting can also be observed above the Fermi level.

\begin{figure}[!b]
\centering
\includegraphics[width=0.95\linewidth]{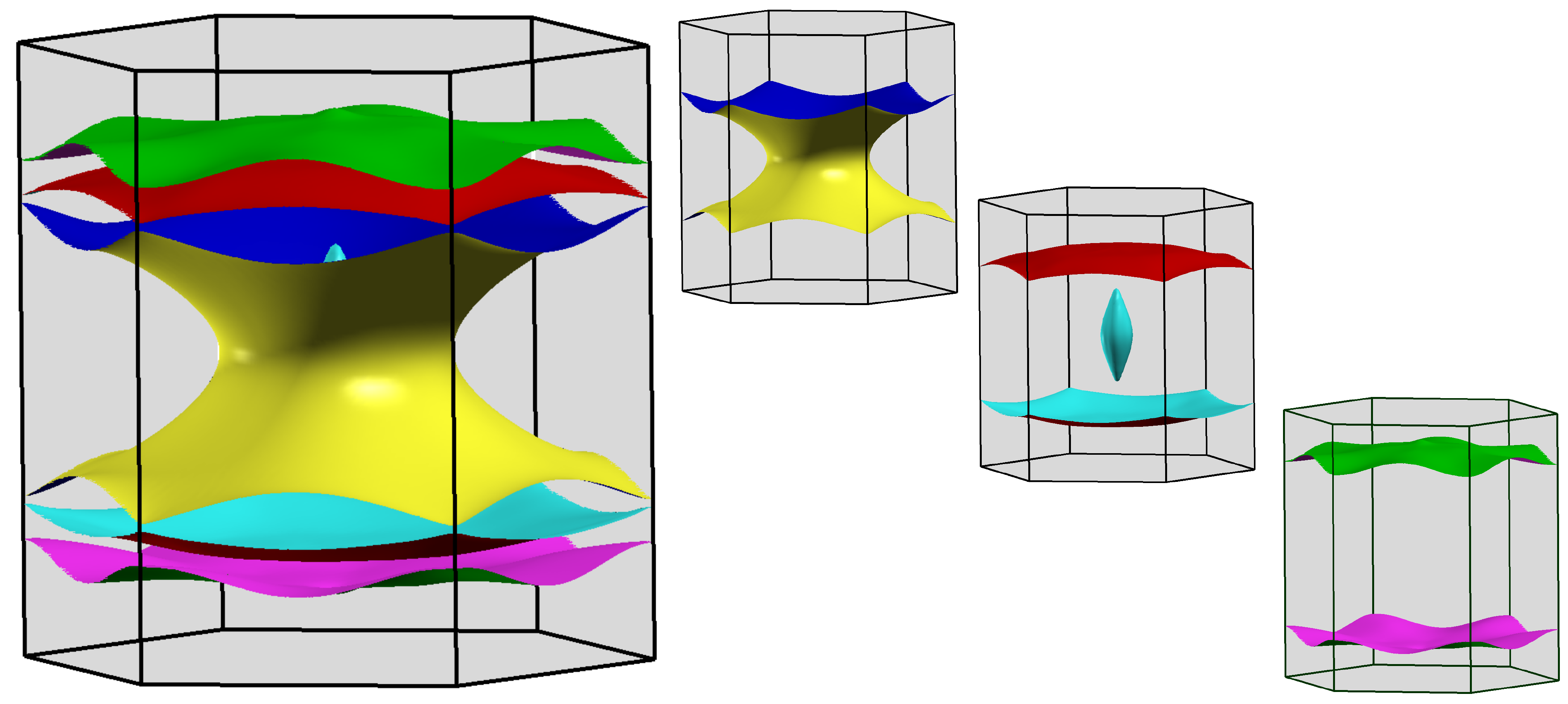}
\caption{
The Fermi surface of RhPb for $P\bar{6}2m$.
The big panel represents the total Fermi surface, while the small panels present the separate Fermi pockets formed by separate bands.
\label{fig.el_fs}
}
\end{figure}

The Fermi surface contains three pockets (Fig.~\ref{fig.el_fs}).
The double-umbrella-like pocket is related to the electronic branch, which exhibits strong dispersion in the $x$-$y$ plane and weak along the $z$ direction, while the ellipsoid pocket exhibits exact 3D electronic behavior.
Similarly, the flat face of two pockets is related to the dispersionless branches in plane $x$-$y$, but relatively strong electronic $k_{z}$-dependence dispersion.
This feature is also reflected in the electronic band structure in the form of a parabolic-like band along $\Gamma$--A path (see Fig.~\ref{fig.el_band}).

\subsection{APRES spectra}

The ARPES experiment was carried out on beamline 21-ID-1 at the National Synchrotron Light Source II, Brookhaven. 
All samples were cleaved in situ under vacuum better than $5 \times 10^{-11}$~Torr. 
Measurements were taken with a synchrotron light source and a Scienta-Omicron DA30 electron analyzer.
The total energy resolution was set to $15$~meV.
The sample stage was maintained at low temperature ($T = 15$~K) throughout the experiment.

\begin{figure}[!t]
\centering
\includegraphics[width=\linewidth]{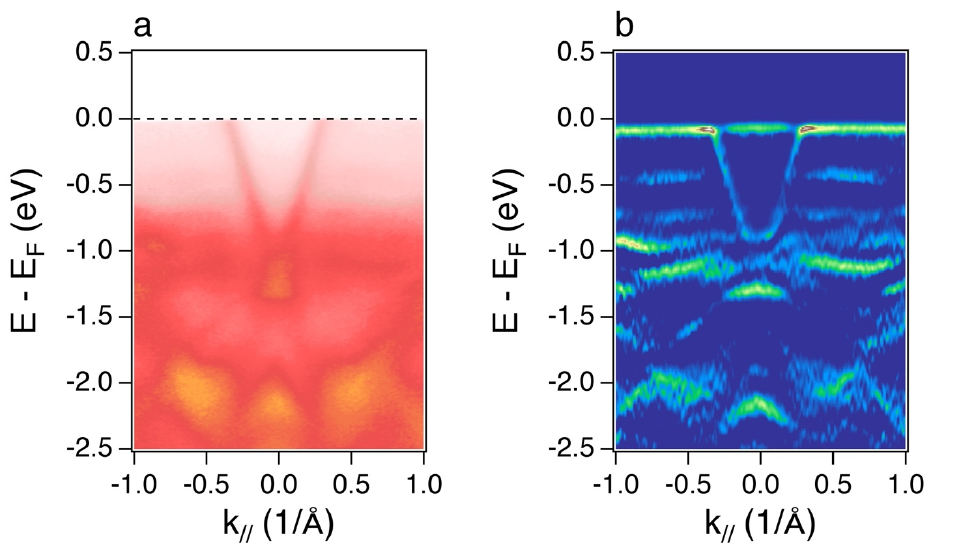}
\caption{
ARPES result for (001) surface along the K-$\Gamma$-K direction. 
ARPES intensity plot (a) and the second derivative of the intensity (b).
The data was collected at incident photon energy $h \nu = 70$~eV, corresponding to $k_{z} = \pi$. 
\label{fig.arpes}
}
\end{figure}

Results of ARPES measurements are presented in Fig.~\ref{fig.arpes}.
Multiple flat-like bands are observed near the binding energy $~1$~eV, consistent with DFT calculations presented in Fig.~\ref{fig.el_band}.
Near the Fermi level, only one parabolic electron-like band is observed at the $\Gamma$ point.

\section{Active modes}
\label{sec_sm.active}

In the case of $P6/mmm$ symmetry, the phonon modes at the $\Gamma$ point can be decomposed into irreducible representation as follows:
\begin{eqnarray}
\Gamma_\text{acoustic} \! \! &=& \! \! A_\text{2u} + \! E_\text{1u} , \\
\nonumber \Gamma_\text{optic} \! \! &=& \! \! 2 A_\text{2u} + \! B_\text{1g} + \! B_\text{1u} + \! B_\text{2u} \\
\nonumber && + \! E_\text{2u} + \! E_\text{2g} + \! 3 E_\text{1u} ,
\end{eqnarray}
where $2 A_\text{2u} + 3 E_\text{1u}$ (acoustic modes not included) are infrared (IR) active, while $E_\text{2g}$ mode (with frequency $3.275$~THz) is Raman active. 
IR active modes are related to the vibration of all atoms in the structure, while Raman active mode only with vibration of atoms in the Wyckoff $2d$ position (i.e., Pb atoms forming honeycomb lattice).
Similarly, in the case $P\bar{6}2m$ we get:
\begin{eqnarray}
\nonumber \Gamma_\text{acoustic} \! &=& \! A_\text{2}'' + \! E' , \\
\Gamma_\text{optic} \! &=& \! A_\text{1}' + \! A_\text{2}' + \! A_\text{1}'' + \! 2 A_\text{2}'' + \! 4 E' + \! E'' ,
\end{eqnarray}
where $2 A_\text{2}'' + 4 E'$ modes are IR active, while $A_\text{1}' + 4 E' + E''$ modes are Raman active.
In the case of this symmetry, both IR and Raman active modes are related to vibrations of all atoms.
Additionally, as we can see, more modes are active in the case of $P\bar{6}2m$ than in $P6/mmm$ symmetry, which could be an additional way to confirm the realized phase.
Nevertheless, the degeneracy of the bands at the $\Gamma$ point is the same in both phases.
Characteristic frequencies and their symmetries can be found in Tab.~\ref{tab.freq}.

\begin{table}[!b]
\caption{
\label{tab.freq}
Characteristic frequencies (THz) and symmetries of the phonon modes at the $\Gamma$ point for RhPb with both symmetries.
}
\begin{ruledtabular}
\begin{tabular}{rlrlrlrl}
\multicolumn{8}{c}{$P6/mmm$} \\
\hline
$-2.590$ & ($B_\text{1u}$) & $0.000$ & ($A_\text{2u}$) & $0.000$ & ($E_\text{1u}$) & $1.675$ & ($A_\text{2u}$) \\
$2.119$ & ($E_\text{1u}$) & $2.519$ & ($E_\text{1u}$) & $3.260$ & ($E_\text{1u}$) & $3.275$ & ($E_\text{2g}$) \\
$3.754$ & ($B_\text{1g}$) & $5.267$ & ($A_\text{2u}$) & $5.355$ & ($E_\text{1u}$) & $6.296$ & ($B_\text{2u}$) \\
\hline
\hline
\multicolumn{8}{c}{$P\bar{6}2m$} \\
\hline
$0.000$ & ($A$) & $0.000$ & ($E$) & $1.777$ & ($A$) & $2.160$ & ($E$) \\
$2.886$ & ($E$) & $3.291$ & ($E$) & $3.296$ & ($A$) & $3.689$ & ($E$) \\
$3.737$ & ($A$) & $5.273$ & ($A$) & $5.507$ & ($E$) & $6.066$ & ($A$) 
\end{tabular}
\end{ruledtabular}
\end{table}


\end{document}